\documentclass[conference,hidelinks,onecolumn]{IEEEtran}
\usepackage{tikz}

\usepackage{setspace}
\usepackage{blindtext}
\usepackage{graphicx}
\usepackage{pgfplots}

\usetikzlibrary{through,calc}

\definecolor{bblue}{HTML}{4F81BD}
\definecolor{rred}{HTML}{C0504D}
\definecolor{ggreen}{HTML}{9BBB59}
\definecolor{ppurple}{HTML}{9F4C7C}

\definecolor{mDarkBrown}{HTML}{604c38}
\definecolor{mDarkTeal}{HTML}{23373b}
\definecolor{mMediumTeal}{HTML}{205A65}

\definecolor{mLightTeal}{HTML}{41a3c3}

\definecolor{mLightBrown}{HTML}{EB811B}
\definecolor{mMediumBrown}{HTML}{C87A2F}

\def\axesradargrid at (#1,#2){\draw[help lines,step=0.5] (0,0) grid (#1,#2);
\draw[->, thin](0,#2/2)--(#1,#2/2);
\node [right] at (#1,#2/2) {$\mathcal{R}$};
\draw[->, thin](#1/2,0)--(#1/2,#2);
\node [above] at (#1/2,#2) {$\mathcal{I}$};}

\def\axesradar at (#1,#2){
\draw[->, thin](0,#2/2)--(#1,#2/2);
\node [right] at (#1,#2/2) {$\mathcal{R}$};
\draw[->, thin](#1/2,0)--(#1/2,#2);
\node [above] at (#1/2,#2) {$\mathcal{I}$};}

\def\pointS at (#1:#2){
\draw [mDarkTeal,  thin,fill=mMediumTeal] (#1:#2) circle [radius=0.04];
\draw [mMediumBrown, thin,dashed] (#1:#2) circle [radius=0.2];}

\usepackage{color,colortbl}

\usepackage{cite}

\ifCLASSINFOpdf
\else
\fi

\usepackage[cmex10]{amsmath}

\usepackage{tikz}

\usepackage{xspace}
\usepackage{graphicx}
\usepackage{epstopdf}
\usepackage{amssymb}
\usepackage[cmex10]{amsmath}
\usepackage[footnotesize]{caption}

\usepackage{subcaption}
\usepackage{cite}

\usepackage{hyperref}

\usepackage{mathtools}
\DeclareMathOperator{\sign}{sign}
\DeclareMathOperator{\supp}{supp}
\DeclarePairedDelimiter\floor{\lfloor}{\rfloor}

\newcommand{\sq}{\vspace{0mm}}
\newcommand{\ie}{\emph{i.e.}, }
\newcommand{\eg}{\emph{e.g.}, }

\newcommand{\ts}{\textstyle}
\newcommand{\bs}{\boldsymbol}
\newcommand{\cl}{\mathcal}
\newcommand{\im}{{\sf i}\mkern1mu}

\usepackage{amsmath,amsfonts,amssymb,amsthm}
\usepackage{graphicx}
\usepackage{xspace}
\usepackage{tikz}
\usepackage{epstopdf}
\usepackage{subcaption}
\newfont{\ninept}{ptmr at 9pt} % for writing captions in 9pt

\usepackage{mathtools}
\makeatletter
\newcommand{\RemoveAlgoNumber}{\renewcommand{\fnum@algocf}{\AlCapSty{\AlCapFnt\algorithmcfname}}}
\newcommand{\RevertAlgoNumber}{\algocf@resetfnum}
\makeatother

\newcommand{\ud}{\mathrm{d}}

\newcommand{\bb}{\mathbb}

\newcommand{\iid}{%
    \ifmmode% math mode
        \mathrm{i.i.d.}%
    \else%
        i.i.d.\@\xspace%
    \fi%
}
\usepackage[colorinlistoftodos]{todonotes}

\usepackage{amsmath,amsfonts,amssymb,amsthm}
\usepackage{graphicx}
\usepackage{xspace}
\usepackage{tikz}
\usepackage{epstopdf}
\usepackage{subcaption}

\usepackage{mathtools}


\usetikzlibrary{external}

\def\htex{1.6in}
\begin{document}

\title{Quantity over Quality: Dithered Quantization\\ for Compressive Radar Systems}
\author{Thomas Feuillen, Chunlei Xu, J\'er\^ome Louveaux, Luc Vandendorpe, Laurent Jacques \\ ICTEAM, UCLouvain, Belgium}

\markboth{Journal of \LaTeX\ Class Files,~Vol.~6, No.~1, January~2007}%
{Shell \MakeLowercase{\textit{et al.}}: Bare Demo of IEEEtran.cls for Journals}

\maketitle

\begin{abstract}
In this paper, we investigate a trade-off between the number of radar observations (or measurements) and their resolution in the context of radar range estimation. To this end, we introduce a novel estimation scheme that can deal with strongly quantized received signals, going as low as 1-bit per signal sample. We leverage for this a dithered quantized compressive sensing framework that can be applied to classic radar processing and hardware. This allows us to remove ambiguous scenarios prohibiting correct range estimation from (undithered) quantized base-band radar signal. Two range estimation algorithms are studied: Projected Back Projection (PBP) and Quantized Iterative Hard Thresholding (QIHT). The effectiveness of the reconstruction methods combined with the dithering strategy is shown through Monte Carlo simulations. Furthermore we show that: \emph{(i)}, in dithered quantization, the accuracy of target range estimation improves when the bit-rate (\ie the total number of measured bits) increases, whereas the accuracy of other undithered schemes saturate in this case; and \emph{(ii)}, for fixed, low bit-rate scenarios, severely quantized dithered schemes exhibit better performances than their full resolution counterparts. These observations are confirmed using real measurements obtained in a controlled environment, demonstrating the feasibility of the method in real ranging applications. 
\end{abstract}

\IEEEpeerreviewmaketitle
\sq
\section{Introduction}
Civilian radar applications such as automotive radar design or the growing fields of smart cities are more and more in need of small form factor and affordable radars \cite{HS09,End10,YONINA}.
As these complex applications often requires the deployment of many radar sensors working in a collaborative mode, the increasing amount of data recorded by these systems challenges both data transmission and processing techniques.  

In this paper, we focus on lightening the acquisition of radar signals; we strongly reduce the resolution (or bit-depth) of each samples collected by a radar sensor without sacrificing accurate depth estimation. More precisely, in order to be processed, the physical voltage signals coming from a radar must be digitized. This quantization process is often forgotten in the system model; high resolutions --- and expensive --- \textit{Analog to Digital Converters} (ADCs) must then be used to mitigate the resulting quantization noise, inducing fairly large bit-rates. We propose to remove this limitation by integrating quantization directly in the signal model, using the framework of \textit{Quantized Compressive Sensing} (QCS)~\cite{BJSK15}. We consider a digitization modeled by a scalar mid-rise uniform quantizer (or lower floor quantizer). Our aim is thus to minimize the impact of lowering the acquisition bit-rate on the quality of the estimated ranges, hence targeting possibly cheaper radar receiver implementations.

The \textit{Compressive Sensing} (CS) theory leverages the low-complexity nature of structured signals (\eg their sparsity, compressibility or low-rankness) to reduce the signal sampling rate at the acquisition~\cite{candes,FR13}. CS shows that, with high probability, one can stably and robustly estimate such signals by collecting a number of random linear measurements driven by the signal ``information-rate", \eg its sparsity level. During the last ten years, many works have considered the association of the radar principles with CS theory: first, to increase a target’s parameter resolution~\cite{HS09}, and later to reduce the number of samples to be processed~\cite{End10}. The survey~\cite{YONINA} describes the reduced sampling rate of different compressive (or sub-Nyquist) radar systems, in comparison with traditional Nyquist sampling schemes, although digitization impact is not covered.

One-bit quantized compressive radar schemes have been studied in, \eg~\cite{MAP,varthres,2block}. One limiting effect is that, as the digitization becomes coarser, ambiguities might appear between different unquantized signals --- and thus different target configurations --- that are digitized to the same bits, rendering the estimation ambiguous. These works, however, failed to address these ambiguities. Our previous work in 1-bit quantization applied to Frequency-Modulated Continuous-Wave radar (FMCW)~\cite{COSERA} showed that these ambiguities do happen in realistic settings and measurements and can be counteracted using a pre-quantization \emph{dither}. Dithering amounts to adding a designed noise on the signal, before quantizer's action, with the goal of attenuating quantization distortions~\cite{ref1, XJ2018}. This procedure is also used in, \eg LIDAR imaging~\cite{ref3} where dithering is implemented in a real set-up by physically varying a time-delay before the acquisition, and was studied for high sampling rate ADCs~\cite{adcdither}.

In this paper we investigate a trade-off between the number of radar observations (or measurements) and their resolutions in the case of an FMCW radar with one transmitting and one receiving antenna. While this setting might seem restrictive as it only considers target range recovery, its setup allows us to perform thorough tests using both simulations and actual radar measurements. To this end, two range estimation algorithms, adapted to quantized radar signal, are used: Projected Back Projection (PBP)~\cite{XJ2018} and Quantized Iterative Hard Thresholding (QIHT)~\cite{kevin}. Compared to~\cite{COSERA}, this work deeply investigates the comparisons between severely quantized and high-resolution measurements constrained to the same bit-rate, \ie between quantity and quality.

Let us summarize the main contributions of this work: \emph{(i)} we show that ambiguities due to the combination of the intrinsic radar Fourier domain with harsh quantization exist and are removed using dithered quantization; \emph{(ii)} we observe that as the number of measurements $M$ grows, non-dithered quantization yields range estimation error (using either PBP or QIHT) that saturates whereas the dithered schemes reach a decaying error when $M$ increases; \emph{(iii)} we show that QIHT provides the best performances at low resolution for harsh bit-rate condition; and \emph{(iv)} we confirm all the above observations on a controlled laboratory set-up using an FMCW radar.

The rest of the paper is structured as follows. In  Sec.~\ref{sec:radar-system-model}, the complete FMCW radar model (\ie its transmission and reception principles) is introduced, as well as a linear inverse problem formulation focused on a Fourier sensing model of the range profile.
Sec.~\ref{sec:quantization} defines the quantization procedure applied on the received radar signal. We then prove that unavoidable ambiguities are induced by this scheme, \ie the existence of distinct received signals (and thus distinct range profiles) whose quantized measurements are identical. A dithered quantizer is then proposed to cancel out these ambiguous situations. Sec.~\ref{sec:algo} describes two algorithms capable to estimate sparse range profiles from quantized observations, namely PBP and QIHT. Finally, we demonstrate the efficiency of our approach through intensive Monte Carlo simulations in Sec.~\ref{sec:simu}, and via real radar measurements in Sec.~\ref{sec:mes_radar}, before concluding in Sec.~\ref{sec:conclusion}.

\textbf{Notations}: Vectors and matrices are denoted with bold symbols. The imaginary unit is $\im = \sqrt{-1}$, $[D] := \{1, \cdots, D\}$ for $D \in \bb N$, ${\rm \bf Id}$ is the identity matrix, $\supp \bs u = \{i : u_i \neq 0\}$ is the support of $\bs u$, $\lfloor \cdot \rfloor$ is the flooring operator, and $|\cl S|$ is the cardinality of a set $\cl S$. $(\cdot)^*$ denotes the complex conjugate and the adjoint operator for scalar and matrices, respectively. For $p \geq 1$, the $\ell_p$-norm of a complex vector $\bs u$ reads $\|\bs u\|_p := (\sum_k |u_k|^p)^{1/p}$, with $\|\bs u\| := \|\bs u\|_2$ and $\|\bs u\|_\infty = \max_k |u_k|$. 
\sq
\section{Radar System Model}
\label{sec:radar-system-model}
We study here an FMCW radar with one transmitting and one receiving antenna. The radar's transmitting antenna emits a signal $s(t)$ modeled as
\begin{equation}
    \ts s(t)=\sqrt{P_t}\exp\big(\im 2\pi (\int_{0}^t f_c(\xi) \ud\xi) +\im\phi_0\big),
    \label{txsig}
\end{equation}
where $P_t$ is the transmitted power, $f_c(t)$ the transmitted frequency pattern, and $\phi_0 \in [0, 2\pi]$ is the initial phase of the oscillator.

\begin{figure}[!t]
\centering
\usetikzlibrary{through,calc}

\definecolor{bblue}{HTML}{4F81BD}
\definecolor{rred}{HTML}{C0504D}
\definecolor{ggreen}{HTML}{9BBB59}
\definecolor{ppurple}{HTML}{9F4C7C}

\definecolor{mDarkBrown}{HTML}{604c38}
\definecolor{mDarkTeal}{HTML}{23373b}
\definecolor{mMediumTeal}{HTML}{205A65}

\definecolor{mLightBrown}{HTML}{EB811B}
\definecolor{mMediumBrown}{HTML}{C87A2F}

\title{Dessin tikz}
\author{Thomas Feuillen}
\date{July 2018}

%%%%%%%%%%%%%%%%%%%%%%%%%%%%%SCHEMABLOC%%%%%%%%%%%%%%%%%%%%

%%%%%%%%%%%%%%%%%%%%%%%%%%%%%%%%%%%%%%%%%%%%%%%%%%%%%%%%%%

\def\axesradargrid at (#1,#2){\draw[help lines,step=0.5] (0,0) grid (#1,#2);
\draw[->, thick](0,#2/2)--(#1,#2/2);
\node [right] at (#1,#2/2) {$\mathcal{R}$};
\draw[->, thick](#1/2,0)--(#1/2,#2);
\node [above] at (#1/2,#2) {$\mathcal{I}$};}

\def\axesradar at (#1,#2){
\draw[->](0,#2/2)--(#1,#2/2);
\node [right] at (#1,#2/2) {$\mathcal{R}$};
\draw[->](#1/2,0)--(#1/2,#2);
\node [above] at (#1/2,#2) {$\mathcal{I}$};}

\def\pointS at (#1:#2){
\draw [mDarkTeal, thick,fill=mMediumTeal] (#1:#2) circle [radius=0.04];
\draw [mMediumBrown, thick,dashed] (#1:#2) circle [radius=0.2];}

\begin{tikzpicture}[thick,scale=6/3, every node/.style={scale=3/3}]
\def\dx{0.32}
%\draw[help lines,step=0.5] (0,0) grid (3,1.5);

\draw[white,fill=mDarkBrown, opacity=0.25] (0,0) rectangle (\dx,1.5) ;

\draw[white,fill=mDarkBrown, opacity=0.25] (1,0) rectangle (1+\dx,1.5) ;

\draw[white,fill=mDarkBrown, opacity=0.25] (2,0) rectangle (2+\dx,1.5) ;

\draw[->, thick] (0,0)--(0,1.5);

\node[above] at (0,1.5) {$f_c(t)$};

\draw[->, thick] (0,0)--(3,0);
\node[right] at (3,0) {$t$};

\draw[thick] (-0.05,0.25)--(0.05,0.25);

\node[left] at (-0.05,0.25) {\footnotesize $f_0$};

\draw[thick] (-0.05,0.25+1)--(0.05,0.25+1);

\node[left] at (-0.05,0.25+1) {\footnotesize $f_0+B$};

\draw[thick] (1,-0.05)--(1,0.05);

\draw[thick] (1+1,-0.05)--(1+1,0.05);
\node[below] at (1,-0.05) {\footnotesize $T$};

\draw[thick] (\dx,-0.05)--(\dx,0.05);
\node[below] at (\dx,-0.05) {\footnotesize $\tau_0$};

\draw[thick,mMediumTeal] (0,0.25)--(1,1.25)--(1,0.25)--(2,1.25)--(2,0.25)--(3,1.25);

\draw[thick,mMediumBrown] (0+\dx,0.25)--(1+\dx,1.25)--(1+\dx,0.25)--(2+\dx,1.25)--(2+\dx,0.25)--(3,1.25-\dx);

\draw[<->,thick,dashed] (0.5+0.25,0.75-\dx+0.25)--(0.5+0.25,0.75+0.25);

\node[above] at (0.58,0.95) {\footnotesize $\frac{B}{T}\tau_0$};

\node[right] at (3,1.25) {\footnotesize \textcolor{mMediumTeal}{$s(t)$}};

\node[right] at (3,1.25-\dx) {\footnotesize \textcolor{mMediumBrown}{$r(t)$}};

\end{tikzpicture}
\caption{The transmitted and received frequency patterns for an FMCW radar in green and orange, respectively.}
\label{fig:FMCW}
\end{figure}
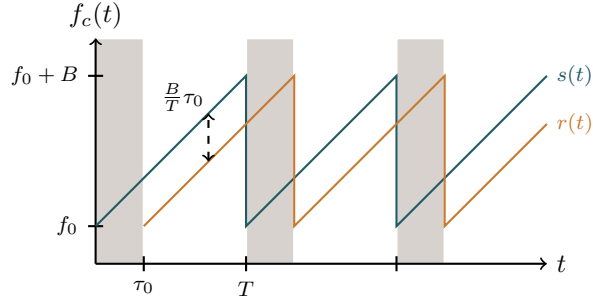

The carrier frequency pattern $f_c(t)$ of an FMCW radar can be characterized as a saw-tooth function (see Fig.~\ref{fig:FMCW}):
\begin{equation}
    \ts f_c(t)=f_0 + B\,(\frac{t}{T} \bmod 1),
    \label{sawtooth}
\end{equation}
with $f_0$ the central frequency, $\bmod$ the modulo operator, $T$ the duration of one ramp, and $B$ the spanned bandwidth. Note that, in practical applications, $B$ is not a design parameter but a constraint imposed by government regulations.

Now considering the received signal model, let us first focus on one static target located at a range $R > 0$ from the receiving antenna. In a noiseless setting, the received signal $r(t)$ is 
\begin{equation}
    r(t) = A s(t-\tau_0),
    \label{delaysig}
\end{equation}
where $A$ is the complex received amplitude, $\tau_0$ is the round-trip delay between the radar and the target and is defined as $\tau_0={2 R}/{\sf c}$, and ${\sf c}$ is the speed of light. For the sake of simplicity, in the rest of our presentation, the complex value $A$ will refer to a global constant amplitude that may change from one line to the other in the description of the reception and demodulation processes.

From~\eqref{txsig} and~\eqref{delaysig}, the received signal is thus: 
\begin{equation}
   \ts r(t)=A\exp\big(\im 2\pi (\int_{0}^{t-\tau_0} f_c(\xi) \ud\xi) + \im \phi_0\big).
    \label{rg}
\end{equation}
   
After coherent base-band demodulation with the transmitted signal~\eqref{txsig}, \ie by replacing $r(t)$ by its multiplication with $s^*(t)$, the expression~\eqref{rg} reduces to 
\begin{equation}
     r(t) = A \exp \big(- \im2\pi \ts \int_{t-\tau_0}^{t} f_c(\xi) \ud\xi  \big).
    \label{rgbb}
\end{equation}
If $f_c$ follows the saw-tooth model~\eqref{sawtooth}, the integral in~\eqref{rgbb} becomes
\begin{align}
\ts \int_{t-\tau_0}^{t} f_c(\xi) \ud \xi &=\ts \int_{t-\tau_0}^{t} (f_0+ \frac{B}{T}\xi) \ud \xi\nonumber\\% =\ts \tau_0 f_0- \frac{B}{2T} \tau_0^2+\tau_0 \frac{B}{T}  t\\
&=\ts \tau_0 f_c(t) - \frac{B}{2T} \tau_0^2.
\label{integral}
\end{align}
Combining~\eqref{rgbb} with~\eqref{integral} allows us to express the received signal $r(t)$ in base-band, \ie we have
\begin{align}
   r(t)& \ts =A \exp\big(-\im 2\pi \tau_0 f_c(t) \big),
\label{freqdif}
\end{align}
where $A$ also encompasses the static phase-shift $- \frac{B}{2T}\tau_0^2$ in~\eqref{integral}.
In words,~\eqref{freqdif} shows that the coherent demodulation expresses the time difference coming from the target as a carrier frequency difference between the transmitted and received signals. This frequency shift linked to the range is represented in Fig.~\ref{fig:FMCW}.
Sampling $r(t)$ at the receiver at a rate $T/N$ for some integer $N$, \ie at time samples $t_m := m (T/N), m\in \mathbb{Z}$, gives
\begin{align}
    %r[m]&=A \exp{\big[-\im 2\pi \ts \frac{2R}{c} \ts \frac{B}{T} t_m  \big]}\\
   % &=A \exp{\im 2 \pi \big[ \ts \frac{2R}{c}f_0 - \ts\frac{2R}{c} \ts\frac{B}{T} T_m-  \frac{2R}{c}f_0 \big]}\\
     r[m]&=A \exp\big(-\im 2\pi f_m \ts\frac{2R}{\sf c} \big),
    \label{sig_sampled}
\end{align}
with $f_m:=f_c(t_m)=f_0+ B\,(\frac{m}{N} \mod 1)$. A single ramp can thus be sampled over at most $N$ time samples, which implicitly determines both the resolution ${\sf c}/(2B)$ and the maximum range $R_{\rm max} := {\sf c} N/(2B)$ at which $R$ can be estimated.

Let us now turn to a multi-target scenario restricted to a purely additive model; all the targets are in a direct line of sight from the radar, without any possible multipath propagation. Taking into account the radar range resolution (${\sf c}/2B$) and $R_{\max}$, we discretize the range domain $(0, R_{\max}]$ with $N$ ranges $\cl R := \{R_n := n ({\sf c}/{2 B}), {1 \leq n \leq N}\}$. A range profile resulting from $K$ targets with ranges in $\cl R$ is expressed as a $K$-sparse vector $\bs a = (a_1,\,\cdots, a_N)^\top$, \ie the amplitude $a_n \neq 0$ if there is a target at the $n^{\rm th}$ range bin $R_n$, and $\|\bs a\|_0 := |\supp \bs a| \le K$. Then, the single target case~\eqref{sig_sampled} generalizes to the multi-target sensing model
\begin{equation}
\ts r[m] =\sum_{n=1}^{N} a_n e^{-\im 2 \pi f_m \frac{2 R_n}{\sf c}} = \sum_{n=1}^{N} a'_n e^{-\im 2 \pi \frac{mn}{N}},
\label{eq:multitargetanalytic}
\end{equation}
where $a'_n = a_n e^{-\im 2\pi n f_0/B}$. In words, each observation $r[m]$ at time $t_m$ amounts to probing the $m^{\rm th}$ frequency of the discrete Fourier transform the range profile $\bs a'=(a'_1, \cdots, a'_N)^\top$. Hereafter, since $\bs a'$ encodes the same range profile than $\bs a$ (up to a modulation), we drop the prime symbol for the sake of simplicity. 
Classically, in a Nyquist sensing scenario, if we collect $N$ samples $\bs r = (r[1],\,\cdots, r[N])^\top$, \eqref{eq:multitargetanalytic} is equivalent to $\bs r =  \bs F^* \bs a$, with $\bs F$ the Fourier matrix ($\ie F_{mn} := \exp(\im 2 \pi \frac{mn}{N})$), and an inverse Fourier transform recovers $\bs a$. For noisy observations, a sampling over multiple ramps --- hence reaching an oversampled sensing model --- yields a robust estimate of $\bs a$.    

In this work, we leverage the sparsity assumption made on $\bs a$ to allow this estimation through severely quantized, possibly oversampled, received signal samples. Without quantization, Compressive Sensing (CS) theory from partial random Fourier sensing matrices shows that, with high probability, we can recover any $K$-sparse vector $\bs a$ from only $M = O(K \log^4 N)$ random samples of~$\bs r$~\cite{FR13}. However, as made clear in Sec.~\ref{sec:quantization}, QCS aims to reduce the impact of signal measurement quantization in signal estimation by possibly increasing the number of measurements beyond $N$; what truly matters in QCS is indeed the total bit-rate $\cl B$ (\ie $M \times$ the bit depth $b$) used to encode the observations~\cite{BJSK15,XJ2018}. 

Consequently, our sensing scheme is determined by sampling the received signal $r(t)$ over a set of $M$ (discrete) time samples $\cl T = \{t'_m: 1 \leq m \leq M\}$ determined as follows. If $M<N$, then $(t'_1, \cdots, t'_M)$ is picked uniformly at random among all possible subset of $M$ time samples of $\{t_m: 1\leq m \leq N\}$. If $M > N$, in an effort to obtain an acquisition time as short as possible, we take then $t'_m = t_m$ for $1\leq m \leq N \lfloor M/N \rfloor$, \ie the first $\lfloor M/N \rfloor$ ramps are fully sampled, and the set of $M' = M - N \lfloor M/N \rfloor$ remaining samples is picked uniformly at random among all possible subset of $M'$ time samples of $\{t_m: N \lfloor M/N \rfloor + 1\leq m \leq N (\lfloor M/N \rfloor +1)\}$, \ie the last ramp is randomly sub-sampled. 

Correspondingly, these time samples are associated with $M$, possibly non-distinct, frequencies $\{f'_m = f_c(t'_m): 1 \leq j \leq M\}$. Finally, if $\Omega$ is a multiset (\ie a set with repeated elements) representing the indices of these frequencies in $[N]$, the final CS model, before quantization, reads
\begin{equation}
    \label{eq:CS-linear-model}
    \bs r = \bs \Phi \bs a = \bs F_{\Omega}^* \bs a,
\end{equation}  
where $\bs \Phi := \bs F^*_{\Omega}$, $\bs F_{\Omega}$ gathers the (possibly repeated) columns of $\bs F$ indexed in $\Omega$, and $\bs r$ follows the sampling of $r(t)$ over $\cl T$. Note that for $M > N$, the addition of a dither ensures that the observations of $r$ over repeated frequencies carry additional information (see Sec.~\ref{sec:quantization}).
\sq
\section{Quantization: Model \& Ambiguity}
\label{sec:quantization}
We select in this work on a uniform $b$-bit scalar quantizer applied componentwise onto complex vectors, separately on the real and the imaginary domains, \ie 
\begin{equation}
    \mathcal{Q}_{b}^\mathbb{C}(\bs r)= \mathcal{Q}_{b}(\Re(\bs r))+\im \mathcal{Q}_{b}(\Im(\bs r)),
    \label{eq:complex-unif-quantizer}
\end{equation}
where $b$ is the number of bits per vector component (\ie the I and Q channels), or \emph{bit depth}. This quantization takes place on the received base-band signal $\bs r$ using ADCs with a resolution of $b$ bits. In~\eqref{eq:complex-unif-quantizer}, $\mathcal{Q}_{b}(\cdot)$ is the standard mid-riser quantizer of quantization step size $\delta>0$~\cite{ref1,XJ2018} 
\begin{equation*}
\ts \mathcal{Q}_{b}(\lambda) := \delta \floor{\frac{\lambda}{\delta}}{} +\frac{\delta}{2},\quad \forall \lambda \in \bb R.
%\label{Qb}    
\end{equation*}
The step size is set to $\delta = \alpha_b \Delta$, where $\Delta$ is the dynamic range of the ADC, \ie its voltage range $[-\Delta,\Delta]$, and $\alpha_b = 2^{1-b}$ ensures that the bit-depth of each sample is $b$. For example, for $b=1$, the ADC is then a simple voltage comparator over its domain, \ie $2\cl Q_1(\cdot)/\Delta \equiv \sign(\cdot)$. This definition assumes that the quantizer is adjusted to the variations of $\bs r$, \ie we must have $\Delta \geq \|\bs r\|_\infty$, with $\Delta$ as small as possible to minimize the quantization distortion which scales like $O(\delta)$. Note that one can also decide to set $\Delta \geq |r[m]|$ only for a significant fraction of indices $m$, \eg if $\|\bs r\|_\infty$ is not bounded. Hereafter, we just assume that $\Delta$ is given.

Let us stress an important limitation of a too direct quantization of the radar sensing model~\eqref{eq:CS-linear-model}: the existence of distinct vectors whose quantized Fourier observations are sent to the same quantized vector, rendering the estimation process ambiguous. This bears similarities with known ambiguities in 1-bit CS with binary matrices~\cite{PV14} and for QCS for multiple antennas and a single target~\cite{COSERA}. We show here that the same effect exists for multiple targets and one receiving antenna.

This ambiguity is explained by the following construction. Given two distinct $n_0, n_1 \in [N]$, we build $\bs a_0 = \bs b_{n_0} e^{-\im \psi_{n_0}}$ and $\bs a_1=\bs  a_0+ \gamma \bs b_{n_1} e^{-\im \psi_{n_1}}$, with $\psi_{n_0}$ and $\psi_{n_1}$ two arbitrary phases in $[-\pi,\pi)$, $0<\gamma<1$, and $\bs b_i \in \{0,1\}^N$ the (canonical) vector whose components are all $0$ but the $i^{\rm th}$ ($i \in [N]$). The signal $\bs a_0$ can be seen as one unit-amplitude target at location $R_{n_0}$, while $\bs a_1$ contains an additional target at $R_{n_1}$ with amplitude $\gamma$. According to the CS model~\eqref{eq:CS-linear-model}, the acquired received signals are $\bs r_0 = \bs \Phi \bs a_0$ and $\bs r_1 = \bs \Phi \bs a_1$, with
\begin{equation}
\begin{array}{rl}
    r_{0}[m]&=e^{-\im\psi_{n_0}} e^{-\im 2\pi \frac{m n_0}{N}},\\
    r_{1}[m]&=e^{-\im\psi_{n_0}}e^{-\im 2\pi \frac{m n_0}{N}}+\gamma e^{-\im\psi_{n_1}}e^{-\im 2\pi \frac{m n_1}{N}},
\end{array}
\label{eq:s0s1}
\end{equation}
Interestingly, there exist parameter values where the quantizer~\eqref{eq:complex-unif-quantizer} sends the two signals to the same quantized vector, \ie for which the ambiguity condition (AC) holds:  
\begin{equation}
\mathcal{Q}_b^{\mathbb{C}}(\bs r_0)=\mathcal{Q}_b^{\mathbb{C}}(\bs r_1). \label{eq:same-quantiz} \tag{AC}
\end{equation}
Consequently, in these cases, while the $\ell_2$-distance $\|\bs a_1 - \bs a_0\| = \gamma$ is non-zero, recovering both $\bs a_1$ and $\bs a_0$ from their identical quantized observations is impossible.
\begin{figure}[!t] 
    \centering
    \null\hfill
    \begin{subfigure}[b]{0.2\textwidth}
       \usetikzlibrary{through,calc}

\definecolor{bblue}{HTML}{4F81BD}
\definecolor{rred}{HTML}{C0504D}
\definecolor{ggreen}{HTML}{9BBB59}
\definecolor{ppurple}{HTML}{9F4C7C}

\definecolor{mDarkBrown}{HTML}{604c38}
\definecolor{mDarkTeal}{HTML}{23373b}
\definecolor{mMediumTeal}{HTML}{205A65}

\definecolor{mLightBrown}{HTML}{EB811B}
\definecolor{mMediumBrown}{HTML}{C87A2F}

\title{Dessin tikz}
\author{Thomas Feuillen}
\date{July 2018}

%%%%%%%%%%%%%%%%%%%%%%%%%%%%%SCHEMABLOC%%%%%%%%%%%%%%%%%%%%

%%%%%%%%%%%%%%%%%%%%%%%%%%%%%%%%%%%%%%%%%%%%%%%%%%%%%%%%%%

%\section{Introduction}

\def\axesradargrid at (#1,#2){\draw[help lines,step=0.5] (0,0) grid (#1,#2);
\draw[->, thick](0,#2/2)--(#1,#2/2);
\node [right] at (#1,#2/2) {$ \Re$};
\draw[->,thick](#1/2,0)--(#1/2,#2);
\node [above] at (#1/2,#2) {$\Im$};}

\def\axesradar at (#1,#2){
\draw[->, thick](0,#2/2)--(#1,#2/2);
\node [right] at (#1,#2/2) {$\Re$};
\draw[->,thick](#1/2,0)--(#1/2,#2);
\node [above] at (#1/2,#2) {$\Im$};}

\def\pointS at (#1:#2){
\draw [mDarkTeal, thick,fill=mMediumTeal] (#1:#2) circle [radius=0.04];
\draw [mMediumBrown,  thick,dashed] (#1:#2) circle [radius=0.2];}

\begin{tikzpicture}[thick,scale=5/3, every node/.style={scale=3/3}]

%\pointS at (10.5:0.5)

\begin{scope}[shift={(1,1)}]

\draw[mMediumTeal, thick] (0,0)--(22.5:0.8) ;

\draw[mMediumBrown,thick] (0,0)--(0.65475,0.4874) ;

\draw[mMediumBrown, thick,dotted] (22.5:0.8)--(0.65475,0.4874) ;
\foreach \i in {0,90,...,270} {{\pointS at (22.5+\i:0.8)};};

\draw [mDarkBrown,  thick,fill=mMediumBrown] (0.65475,0.4874) circle [radius=0.02];

\node[ right] at (22.5:0.8)  {\footnotesize \textcolor{mMediumTeal}{$r_0[m]$}};

\node[above right] at (0.65475/2+0.15,0.4874/2+0.17)  {\footnotesize \textcolor{mMediumBrown}{$r_1[m]$}};

%\node[above left] at (10,90) {(a)};

\end{scope}

\axesradar at (2,2)
\end{tikzpicture}
        \caption{}
        \label{fig:counter}
    \end{subfigure}\hfill
    \begin{subfigure}[b]{0.2\textwidth}
        \usetikzlibrary{through,calc}

\definecolor{bblue}{HTML}{4F81BD}
\definecolor{rred}{HTML}{C0504D}
\definecolor{ggreen}{HTML}{9BBB59}
\definecolor{ppurple}{HTML}{9F4C7C}

\definecolor{mDarkBrown}{HTML}{604c38}
\definecolor{mDarkTeal}{HTML}{23373b}
\definecolor{mMediumTeal}{HTML}{205A65}

\definecolor{mLightBrown}{HTML}{EB811B}
\definecolor{mMediumBrown}{HTML}{C87A2F}

\title{Dessin tikz}
\author{Thomas Feuillen}
\date{July 2018}

%%%%%%%%%%%%%%%%%%%%%%%%%%%%%SCHEMABLOC%%%%%%%%%%%%%%%%%%%%

%%%%%%%%%%%%%%%%%%%%%%%%%%%%%%%%%%%%%%%%%%%%%%%%%%%%%%%%%%

\def\axesradargrid at (#1,#2){\draw[help lines,step=0.5] (0,0) grid (#1,#2);
\draw[->, thick](0,#2/2)--(#1,#2/2);
\node [right] at (#1,#2/2) {$ \Re$};
\draw[->,thick](#1/2,0)--(#1/2,#2);
\node [above] at (#1/2,#2) {$\Im$};}

\def\axesradar at (#1,#2){
\draw[->, thick](0,#2/2)--(#1,#2/2);
\node [right] at (#1,#2/2) {$ \Re$};
\draw[->, thick](#1/2,0)--(#1/2,#2);
\node [above] at (#1/2,#2) {$\Im$};}

\def\pointS at (#1:#2){
\draw [mDarkTeal, thick,fill=mMediumTeal] (#1:#2) circle [radius=0.04];
\draw [mMediumBrown, thick,dashed] (#1:#2) circle [radius=0.2];}

\begin{tikzpicture}[thick,scale=5/3, every node/.style={scale=3/3}]

%\pointS at (10.5:0.5)

\begin{scope}[shift={(1,1)}]

\draw [white,  thick,fill=mDarkBrown,opacity=0.25] (45:0.8) circle [radius=0.3];
\draw [white, thick,fill=white] (45:0.8) circle [radius=0.1];

\draw [mDarkBrown,dashed,  thick] (0.4811,0.7469) circle [radius=0.1];

\draw[mDarkBrown, thick,dotted] (0.4811,0.7469)--(0.4811-0.1,0.7469) ;
\draw[mMediumTeal, thick] (0,0)--(45:0.8) ;
\draw[mMediumBrown, thick] (0,0)--(0.4811,0.7469) ;
\draw[mMediumBrown, thick,dotted] (45:0.8)--(0.4811,0.7469) ;
\foreach \i in {0} {{\pointS at (45+\i:0.8)};};

\draw [mDarkBrown,  thick,fill=mMediumBrown] (0.4811,0.7469) circle [radius=0.02];

\draw [mDarkBrown,  thick,fill=mDarkBrown] (0.4811-0.1,0.7469) circle [radius=0.01];

\draw[mDarkBrown, thick] (0,0)--(0.4811-0.1,0.7469) ;

\node[left ] at (0.4811-0.13,0.7469-0.07) {\footnotesize \textcolor{mDarkBrown}{$r_2$}};

\node[ right] at (40:0.90/3.2)  {\footnotesize \textcolor{mMediumTeal}{$r_0$}};

\node[above right] at (0-0.05,0.7469/2)  {\footnotesize \textcolor{mMediumBrown}{$r_1$}};

\end{scope}

\axesradar at (2,2)
\end{tikzpicture}
        \caption{}
        \label{fig:recurse}
    \end{subfigure}\hfill\null
    \caption{\ninept (a) Graphical representation of $\bs r_0$ and $\bs r_1$ and the domain on which $\bs r_1$ lies. (b) Extension to 3 targets, where the domain to consider for the verification of~\eqref{eq:same-quantiz} is enlarged.}
    \label{K1}
\end{figure}
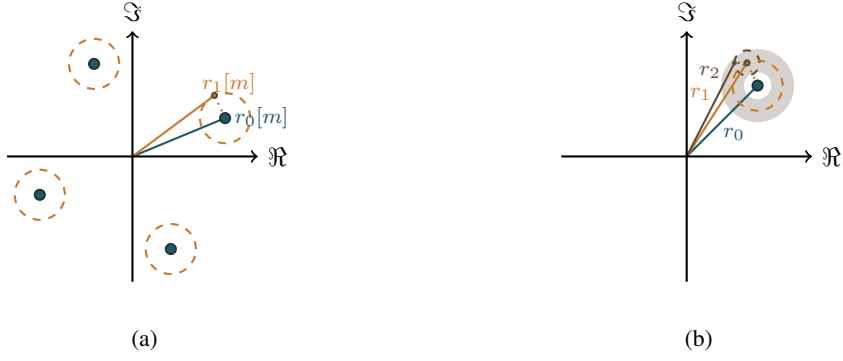
Let us study when~\eqref{eq:same-quantiz} occurs for 1-bit quantization ($b=1$), \ie $\mathcal{Q}^{\mathbb{C}}_1(\cdot) \propto \sign\left(\Re( \cdot)\right)+\im \sign\left(\Im(\cdot)\right)$. In this case,~\eqref{eq:same-quantiz} involves that $r_0[m]$ and $r_1[m]$ are always in the same quadrant of the complex plane $\bb C$ for all $m$. Since from~\eqref{eq:s0s1} $r_1[m]$ lies on a circle of center $r_0[m]$ and radius $\gamma$ in $\bb C$, regardless of the values of $\psi_{n_1}$ or $R_{n_1}$ (see Fig.~\ref{fig:counter}),~\eqref{eq:same-quantiz} holds if  
\begin{equation}
    \min_{m}\, \min(|\Re(r_0[m])|,\, |\Im(r_0[m])|)\ >\ \gamma. \label{cond}
\end{equation}
As $r_0[m] = e^{-\im(\psi_0 + 2\pi \frac{n_0 m}{N})}$,~\eqref{cond} shows a clear dependency between the parameters $\psi_0$, $N$, $M$, and $n_0$ for two quantized vectors to be indistinguishable. For instance, if $n_0 = N/4$, then we just need $\gamma < \min(| \sin \psi_0|,| \cos \psi_0|)$ for~\eqref{eq:same-quantiz} to hold for any values of $\psi_1$ and $n_1$ (see Fig.~\ref{fig:counter}). Similar examples can be constructed for other values of $n_0$, as well as with multiple targets, with then more restriction on the amplitudes of the additional targets as suggested in Fig.~\ref{fig:recurse}. For $b>1$, there also exist vectors satisfying~\eqref{eq:same-quantiz}, but their $\ell_2$-distance must decay if $b$ increases since $\cl Q_b$ splits $\bb C$ into square cells of size $2^{1-b} \Delta$. Therefore, if an algorithm wrongly estimates $\bs r_1$ with the value of $\bs r_0$, its error decays as $2^{-b}$ if $b$ increases, but this error is not ensured to decay if $M$ increases. 

In this work, we stress that the previous ambiguities can be removed by voluntary introducing randomness in the quantization, \ie by inserting a random dither in the quantizer input. Consequently, one can design algorithms whose estimation error of range profile decay as $M$ increases. While dithered quantization is a well-known strategy to improve signal estimation techniques (see, \eg~\cite{ref1,ref2,ref3}), its use in quantized compressive sensing is recent and we follow here the approach of~\cite{XJ2018}.

Given a range profile $\bs a \in \bb C^N$, our dithered QCS sensing model is thus defined by
\begin{equation}
    \bs y = \mathcal{A}_b(\bs a) :=\mathcal{Q}_b^{\mathbb{C}}(\bs \Phi \bs a + \bs \xi),
    \label{eq:ditheredquant}
\end{equation}
where $\bs \xi \in \bb C^M$ is a complex dither defined as $\xi_i=\xi_i^\Re+\im\xi_i^\Im$, with $\xi_i^\Re, \xi_i^\Im \sim \mathcal{U}(-\frac{\delta}{2},\frac{\delta}{2})$. This dither induces more diversity in the quantized measurements, especially for $M>N$. Moreover, $\bb E_\xi \mathcal{A}_b(\bs a) = \bs \Phi \bs a$, \ie the dither cancels out the quantization error in expectation, or, equivalently, if $M$ is large~\cite{XJ2018}. Note that this also changes the dynamic range of the signal before quantization, \ie we must adapt the range $\Delta \geq\|\bs r \|_\infty+\frac{\delta}{2}$.
\sq
\section{Reconstruction algorithm}
\label{sec:algo}
To reconstruct the range profile $\bs a$ from the quantized measurements $\bs y$, two algorithms are studied. The first is Projected Back Projection (PBP) and is defined as follow:

\begin{equation*}
    \ts \hat{\bs a} = \cl H_K \big(\frac{1}{M} \bs \Phi^* \bs y \big),
\end{equation*}
where $K$ is the range profile sparsity, assumed known a priori, $\mathcal{H}_K$ is the hard-thresholding operator setting all the components of its vector input to zero but those with the $K$ largest amplitudes. 

The advantages of PBP are threefold. First, its complexity is $O(N \log N)$ since $\bs \Phi^*$ only requires the computation of an inverse FFT applied on a zero-padding\footnote{In this sense, PBP is similar to a \textit{Maximum Likelihood Estimator}.} of $\bs y$ from $\Omega$ to $[N]$ (or $[\rho N]$ for $\rho = O(1)$ ramps) and $\cl H_K$ involves a vector component ordering of $O(N \log N)$ computations. Second, as a function of $\bs y$, PBP does not explicitly invoke the dither $\bs \xi$; its implementation only requires the knowledge of $\bs \Phi$, \ie of $\Omega$. Finally, in the context of dithered QCS, PBP enjoys of a reconstruction error that decays when $M$ increases for all sensing matrices $\bs \Phi$ respecting with high probability the restricted isometry property (RIP)~\cite{XJ2018}, such as for the random partial Fourier matrix in~\eqref{eq:CS-linear-model}. For a sparse range profile $\bs a$, the reconstruction guarantees is
\begin{equation}
    \|\bs a-\hat{\bs a}\| = \mathcal{O}(M^{-\frac{1}{2}}).
    \label{PBPguarantee}
\end{equation}
In other words, compared to the undithered context, no counterexamples exist that would make this error stagnate when $M$ is increased.

Note that~\eqref{PBPguarantee} is a root-mean-square error bound for the estimation of $\bs a$. In this work, our interest is, however, to characterize the range recovery of target, \ie the support of~$\bs a$. Interestingly, since $\|\bs a-\hat{\bs a}\|_\infty \leq \|\bs a-\hat{\bs a}\|$, if $\bs a$ is $K$-sparse, with $K$ given, and if we know that $\min\{|a_i|: i \in \supp \bs a\} > \eta$ for some $\eta > 0$, then, one can expect that 
\begin{equation}
\label{eq:link-M-min-target-ampl-and-support-recov}
M \geq C/\eta^2\quad \Rightarrow\quad \supp \hat{\bs a} = \supp \bs a,
\end{equation}
for some $C>0$. Indeed, support recovery is ensured if $|\hat{a}_i| > |\hat{a}_j|$ for all $i \in \supp \bs a$ and all $j \in [N]\setminus \{\supp \bs a\}$, which is achieved if $|a_i| - |\hat{a}_i - a_i| > |\hat{a}_j - a_j|$. This holds if $|a_i| > \eta > 2 \|\bs a - \hat{\bs a}\|_\infty = O(M^{-1/2})$, or if $M\geq C/\eta^2$.

While requiring a single iteration, PBP does not ensure that its estimate $\hat{\bs a}$ is \emph{consistent} with $\bs y$, \ie $\cl A_b(\hat{\bs a}) \neq \bs y = \cl A_b({\bs a})$; the quantized sensing model is thus not fully exploited while estimating $\bs a$ from $\bs y$. To solve this situation,~\cite{kevin} has proposed the Quantized Iterative Hard Thresholding (QIHT) algorithm, \ie a variant of the Iterative Hard Thresholding (IHT)~\cite{BD2009} and of the Binary IHT~\cite{JLBB13}, iteratively enforcing both consistency and sparsity of a signal estimate. QIHT is defined by
\begin{equation}
    \hat{\bs a}^{j+1} = \ts \mathcal{H}_K \big[ \hat{\bs a}^j+ \frac{\mu}{M} \bs \Phi^* \big(\bs y - \mathcal{A}_b(\hat{\bs a}^j ) \big)\big],
    \label{eq:QIHT}
\end{equation}
where $j$ is the iteration index, $\mu$ is a step size parameter, and $\hat{\bs a}^0$ is the PBP estimate. Compared to PBP, this algorithm is not ensured to converge. However, numerically, QIHT often provides a sparse and consistent estimate. If this happens at the $J^{\rm th}$ iteration, \ie $\bs y - \cl A_b(\hat{\bs a}^J) = \bs 0$, and if $\bs \Phi$ is a random Gaussian matrix, the QIHT estimate $\hat{\bs a} = \hat{\bs a}^J$ reaches an error $\|\bs a -\hat{\bs a}\| = O(1/M)$~\cite{JL16}. Consequently, we decide to also investigate the efficiency of QIHT for the radar sensing model~\eqref{eq:ditheredquant}.

While QIHT has more to offer in terms of reconstruction by enforcing the consistency, one must also note that knowing the dither at the reconstruction, as imposed by the computation of $\mathcal{A}_b$ in~\eqref{eq:QIHT}, will impact the physical implementation of the system. Indeed PBP could use analogical random noise source such as a noise diode~\cite{adcdither}, whereas QIHT would require a more advanced implementation.
\sq
\section{Numerical Results}
\label{sec:simu}
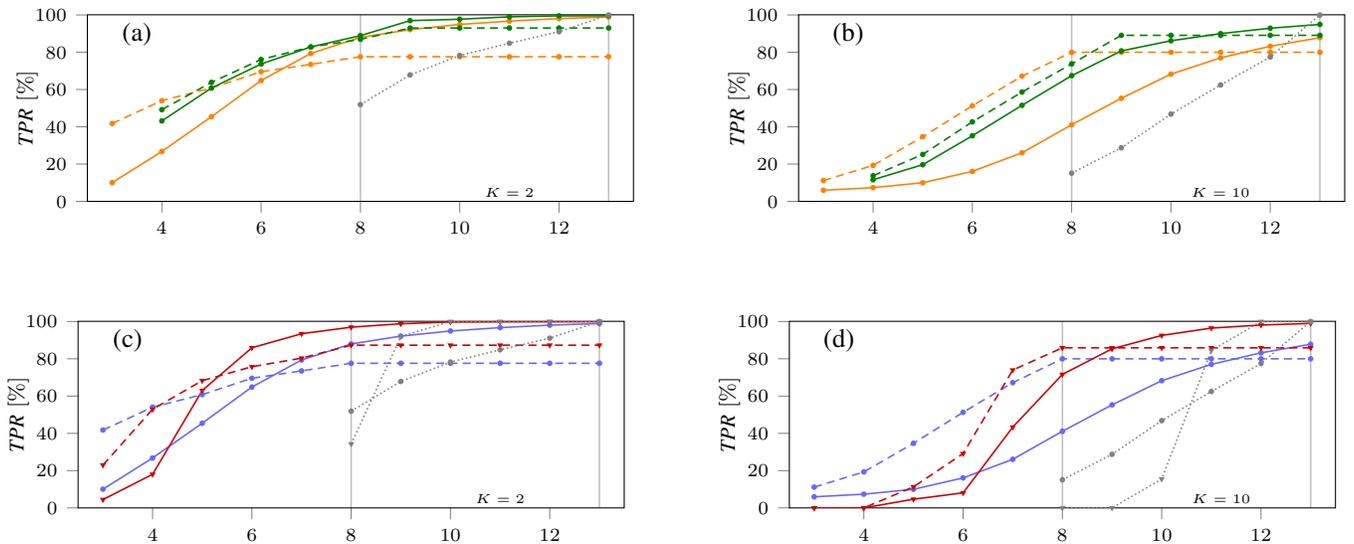
\begin{figure*}[!ht] 
    \centering
    \captionsetup[subfigure]{labelformat=empty}
    \null\ \begin{subfigure}[b]{0.48\columnwidth}
       % This file was created by matplotlib2tikz v0.6.17.
\begin{tikzpicture}

%\definecolor{color1}{rgb}{0.415686274509804,0.352941176470588,0.803921568627451}
%\definecolor{color0}{rgb}{0.117647058823529,0.564705882352941,1}
%\definecolor{color3}{rgb}{0.862745098039216,0.0784313725490196,0.235294117647059}
%\definecolor{color2}{rgb}{0.580392156862745,0,0.827450980392157}

\def\sizep{0.75}

\begin{axis}[
width=1.\columnwidth,
height=\htex,
ylabel={\footnotesize \textit{TPR} $[\%]$},
xmin=2.5, xmax=13.5,
ymin=0, ymax=100.1,
xtick={0,2,...,12,14},
ytick={0,20,...,80,100},
ylabel near ticks,
xlabel near ticks,
x grid style={lightgray!92.02614379084967!black},
y grid style={lightgray!92.02614379084967!black},
ylabel shift = -8 pt,
tick align=outside,
tick pos=left,
ticklabel style = {font=\fontsize{7}{1}\selectfont}
]

\node at (10,90) {\fontsize{10}{1}\selectfont (a)};
\node at (85,05) {\fontsize{4}{1}\selectfont $K=2$};

\addplot [semithick,orange, mark=*, mark size=\sizep, mark options={solid}]
table {%
3 10.0380589914367
4 26.7840152235966
5 45.4329210275928
6 64.8430066603235
7 79.3529971455756
8 87.9638439581351
9 92.1503330161751
10 94.9096098953378
11 96.7174119885823
12 98.0494766888677
13 98.9058039961941
};
\addplot [semithick, orange, densely dashed, mark=*, mark size=\sizep, mark options={solid}]
table {%
3 41.7697431018078
4 54.0437678401522
5 60.7992388201713
6 69.5528068506185
7 73.5014272121789
8 77.592768791627
9 77.592768791627
10 77.592768791627
11 77.592768791627
12 77.592768791627
13 77.592768791627
};

\addplot [semithick, green!50!black, mark=*, mark size=\sizep, mark options={solid}]
table {%
4 43.1493815413891
5 60.7992388201713
6 73.6917221693625
7 82.825880114177
8 88.9153187440533
9 96.9552806850618
10 97.7164605137964
11 99.0009514747859
12 99.5242626070409
13 99.5718363463368
14 nan
};
\addplot [semithick, green!50!black, densely dashed, mark=*, mark size=\sizep, mark options={solid}]
table {%
4 49.1912464319695
5 63.8439581351094
6 76.1179828734539
7 82.825880114177
8 86.9172216936251
9 93.0066603235014
10 93.0066603235014
11 93.0066603235014
12 93.0066603235014
13 93.0066603235014
14 nan
};

\addplot [semithick, lightgray!66.92810457516339!black, densely dotted, mark=*, mark size=\sizep, mark options={solid}]
table {%
8 51.9029495718363
9 67.8401522359657
10 78.2588011417697
11 84.8239771646051
12 91.0561370123692
13 100
14 nan
15 nan
16 nan
17 nan
18 nan
};
\addplot [semithick, black, nearly transparent, forget plot]
table {%
8 0
8 100
};
\addplot [semithick, black, nearly transparent, forget plot]
table {%
13 0
13 100
};

\end{axis}

\end{tikzpicture}
        \caption{}
        \label{fig:PBPtexK2}
    \end{subfigure}\hfill
    \begin{subfigure}[b]{0.48\columnwidth}
        % This file was created by matplotlib2tikz v0.6.17.
\begin{tikzpicture}

\definecolor{color1}{rgb}{0.415686274509804,0.352941176470588,0.803921568627451}
\definecolor{color0}{rgb}{0.117647058823529,0.564705882352941,1}
\definecolor{color3}{rgb}{0.862745098039216,0.0784313725490196,0.235294117647059}
\definecolor{color2}{rgb}{0.580392156862745,0,0.827450980392157}

\def\sizep{0.75}

\begin{axis}[
width=1.\columnwidth,
height=\htex,
ylabel={\footnotesize \textit{TPR} $[\%]$},
xmin=2.5, xmax=13.5,
ymin=0, ymax=100.1,
xtick={0,2,...,12,14},
ytick={0,20,...,80,100},
ylabel near ticks,
xlabel near ticks,
x grid style={lightgray!92.02614379084967!black},
y grid style={lightgray!92.02614379084967!black},
ylabel shift = -8 pt,
tick align=outside,
tick pos=left,
ticklabel style = {font=\fontsize{7}{1}\selectfont}
]

\node at (10,90) {\fontsize{10}{1}\selectfont (b)};
\node at (85,05) {\fontsize{4}{1}\selectfont $K=10$};

\addplot [semithick, orange, mark=*, mark size=\sizep, mark options={solid}]
table {%
3 5.95623215984777
4 7.36441484300666
5 9.98097050428164
6 16.0894386298763
7 26.0704091341579
8 41.1037107516651
9 55.290199809705
10 68.2492863939106
11 77.0123691722169
12 83.2064700285442
13 87.8116079923882
};
\addplot [semithick, orange, densely dashed, mark=*, mark size=\sizep, mark options={solid}]
table {%
3 11.151284490961
4 19.2959086584206
5 34.6717411988582
6 51.2844909609895
7 67.1836346336822
8 79.9904852521408
9 79.9904852521408
10 79.9904852521408
11 79.9904852521408
12 79.9904852521408
13 79.9904852521408
};

\addplot [semithick, green!50!black, mark=*, mark size=\sizep, mark options={solid}]
table {%
4 11.5984776403425
5 19.6574690770695
6 35.185537583254
7 51.5223596574691
8 67.4215033301617
9 80.7231208372978
10 86.1655566127498
11 89.9904852521408
12 92.9019980970504
13 94.919124643197
14 nan
};
\addplot [semithick, green!50!black, densely dashed, mark=*, mark size=\sizep, mark options={solid}]
table {%
4 13.7392959086584
5 25.1760228353949
6 42.6736441484301
7 58.6489058039962
8 73.7583254043768
9 89.0770694576594
10 89.0770694576594
11 89.0770694576594
12 89.0770694576594
13 89.0770694576594
14 nan
};

\addplot [semithick, lightgray!66.92810457516339!black, densely dotted, mark=*, mark size=\sizep, mark options={solid}]
table {%
8 15.1094196003806
9 28.744053282588
10 46.8125594671741
11 62.4548049476689
12 77.4405328258801
13 100
14 nan
15 nan
16 nan
17 nan
18 nan
};
\addplot [semithick, black, nearly transparent, forget plot]
table {%
8 0
8 100
};
\addplot [semithick, black, nearly transparent, forget plot]
table {%
13 0
13 100
};

\end{axis}
\end{tikzpicture}
        \caption{}
        \label{fig:PBPtexK10}
    \end{subfigure}\hfill
    \begin{subfigure}[b]{0.48\columnwidth}
       % This file was created by matplotlib2tikz v0.6.17.
\begin{tikzpicture}

\definecolor{color0}{rgb}{0.862745098039216,0.0784313725490196,0.235294117647059}
\def\sizep{0.75}
\begin{axis}[
xmin=2.5, xmax=13.5,
ymin=0, ymax=100.1,
xtick={0,2,...,12,14},
ytick={0,20,...,80,100},
width=1.\columnwidth,
height=\htex,
ylabel near ticks,
xlabel near ticks,
ylabel={\footnotesize \textit{TPR} $[\%]$},
ylabel shift = -8 pt,
tick align=outside,
tick pos=left,
ticklabel style = {font=\fontsize{7}{1}\selectfont},
x grid style={lightgray!92.02614379084967!black},
y grid style={lightgray!92.02614379084967!black}
]
 
\node at (10,90) {\fontsize{10}{1}\selectfont (c)};
\node at (85,05) {\fontsize{4}{1}\selectfont $K=2$};

\addplot [semithick, blue!60!white, mark=*, mark size=\sizep, mark options={solid}]
table {%
3 10.0380589914367
4 26.7840152235966
5 45.4329210275928
6 64.8430066603235
7 79.3529971455756
8 87.9638439581351
9 92.1503330161751
10 94.9096098953378
11 96.7174119885823
12 98.0494766888677
13 98.9058039961941
};
\addplot [semithick, blue!60!white, densely dashed, mark=*, mark size=\sizep, mark options={solid}]
table {%
3 41.7697431018078
4 54.0437678401522
5 60.7992388201713
6 69.5528068506185
7 73.5014272121789
8 77.592768791627
9 77.592768791627
10 77.592768791627
11 77.592768791627
12 77.592768791627
13 77.592768791627
};

\addplot [semithick, red!80!black, mark=triangle*, mark size=\sizep, mark options={solid,rotate=180}]
table {%
3 4.47193149381541
4 18.0304471931494
5 62.9876308277831
6 85.918173168411
7 93.4348239771646
8 96.9552806850618
9 98.8106565176023
10 99.7145575642246
11 99.8097050428164
12 99.8572787821123
13 99.9048525214082
};
\addplot [semithick, red!80!black, densely dashed, mark=triangle*, mark size=\sizep, mark options={solid,rotate=180}]
table {%
3 22.9781160799239
4 52.9019980970504
5 68.2683158896289
6 75.7373929590866
7 80.3520456707897
8 87.2978116079924
9 87.2978116079924
10 87.2978116079924
11 87.2978116079924
12 87.2978116079924
13 87.2978116079924
};

\addplot [semithick, lightgray!66.92810457516339!black, densely dotted, mark=*, mark size=\sizep, mark options={solid}]
table {%
8 51.9029495718363
9 67.8401522359657
10 78.2588011417697
11 84.8239771646051
12 91.0561370123692
13 100
14 nan
15 nan
16 nan
17 nan
18 nan
};
\addplot [semithick, black, nearly transparent, forget plot]
table {%
8 0
8 100
};
\addplot [semithick, black, nearly transparent, forget plot]
table {%
13 0
13 100
};
\addplot [semithick, lightgray!66.92810457516339!black, densely dotted, mark=triangle*, mark size=\sizep, mark options={solid,rotate=180}]
table {%
8 34.3006660323501
9 91.9600380589914
10 100
11 100
12 100
13 100
14 nan
15 nan
16 nan
17 nan
18 nan
};

\end{axis}

\end{tikzpicture}
        \caption{}
        \label{fig:compare12bitK2}
    \end{subfigure}\hfill
    \begin{subfigure}[b]{0.48\columnwidth}
        % This file was created by matplotlib2tikz v0.6.17.
\begin{tikzpicture}

\definecolor{color0}{rgb}{0.862745098039216,0.0784313725490196,0.235294117647059}
\def\sizep{0.75}
\begin{axis}[
xmin=2.5, xmax=13.5,
ymin=0, ymax=100.1,
xtick={0,2,...,12,14},
ytick={0,20,...,80,100},
width=1.\columnwidth,
height=\htex,
ylabel near ticks,
xlabel near ticks,
ylabel={\footnotesize \textit{TPR} $[\%]$},
ylabel shift = -8 pt,
tick align=outside,
tick pos=left,
ticklabel style = {font=\fontsize{7}{1}\selectfont},
x grid style={lightgray!92.02614379084967!black},
y grid style={lightgray!92.02614379084967!black}
]

\node at (10,90) {\fontsize{10}{1}\selectfont (d)};
\node at (87,05) {\fontsize{4}{1}\selectfont $K=10$};

\addplot [semithick, blue!60!white, mark=*, mark size=\sizep, mark options={solid}]
table {%
3 5.95623215984777
4 7.36441484300666
5 9.98097050428164
6 16.0894386298763
7 26.0704091341579
8 41.1037107516651
9 55.290199809705
10 68.2492863939106
11 77.0123691722169
12 83.2064700285442
13 87.8116079923882
};
\addplot [semithick, blue!60!white, densely dashed, mark=*, mark size=\sizep, mark options={solid}]
table {%
3 11.151284490961
4 19.2959086584206
5 34.6717411988582
6 51.2844909609895
7 67.1836346336822
8 79.9904852521408
9 79.9904852521408
10 79.9904852521408
11 79.9904852521408
12 79.9904852521408
13 79.9904852521408
};
\addplot [semithick, black, nearly transparent, forget plot]
table {%
8 0
8 100
};
\addplot [semithick, black, nearly transparent, forget plot]
table {%
13 0
13 100
};
\addplot [semithick, red!80!black, mark=triangle*, mark size=\sizep, mark options={solid,rotate=180}]
table {%
3 0
4 0
5 4.66222645099905
6 8.07802093244529
7 43.3396764985728
8 71.6270218839201
9 85.3472882968601
10 92.5309229305423
11 96.4700285442436
12 98.116079923882
13 98.9819219790676
};
\addplot [semithick, red!80!black, densely dashed, mark=triangle*, mark size=\sizep, mark options={solid,rotate=180}]
table {%
3 0
4 0.00951474785918173
5 11.2940057088487
6 29.1436726926736
7 73.9676498572788
8 85.8801141769743
9 85.8801141769743
10 85.8801141769743
11 85.8801141769743
12 85.8801141769743
13 85.8801141769743
};

\addplot [semithick, lightgray!66.92810457516339!black, densely dotted, mark=*, mark size=\sizep, mark options={solid}]
table {%
8 15.1094196003806
9 28.744053282588
10 46.8125594671741
11 62.4548049476689
12 77.4405328258801
13 100
14 nan
15 nan
16 nan
17 nan
18 nan
};

\addplot [semithick, lightgray!66.92810457516339!black, densely dotted, mark=triangle*, mark size=\sizep, mark options={solid,rotate=180}]
table {%
8 0
9 0
10 15.4900095147479
11 84.4433872502379
12 100
13 100
14 nan
15 nan
16 nan
17 nan
18 nan
};

\end{axis}
\end{tikzpicture}
        \caption{}
        \label{fig:compare12bitK10}
    \end{subfigure}\ \null

    \caption{[best viewed in color] (a) and (b): TPR vs $\log_2 \cl B$ for PBP; (c) and (d): Comparison between PBP (disks) and QIHT (triangles) in function of $\log_2 \cl B$. In all figures, solid, dashed and dotted curves stand for dithered, undithered and unquantized schemes, respectively. The first (second) gray vertical line represents a bit-rate of $2^8$ ($2^{13}$) bits corresponding to $M=256$ ($M=8192$) for $1$-bit and $M=16$ ($M=256$) for no quantization. In (a) and (b), the resolution is represented by colors, orange for $1$-bit, green for $2$-bits and gray in absence of quantization. In (c) and (d) blue stands for $1$-bit PBP, red for $1$-bit QIHT and gray for no quantization. Figures (a,c) and (b,d) are for $K=2$ and $K=10$, respectively.}
\end{figure*}
We here challenge the possibility of recovering sparse range profiles from quantized radar observations, \ie from measurements associated with the dithered QCS model~\eqref{eq:ditheredquant}. To this aim, we present the result of extensive Monte Carlo (MC) simulations for various parameters of our setup: we have set the sparsity level $K$ --- the number of targets --- in $[2, 10]$, a total bit-rate $\cl B = bM$ in $[2^3, 2^{13}]$ with measurement number $M$ in $[2^3,2^{13}]$ and a bit depth $b \in [1,32]$, $N=256$. Concerning QIHT, we have set $\mu = 1$ and a total number of iterations between $20$ and $100 K$, with an early stop if, either, the \emph{consistency} ${M^{-1} \sum_k (\cl A_b(\hat{\bs a}^j)_k\!=\!y_k)}$ between $\hat{\bs a}^j$ and $\bs a$ exceeds $95\%$, or if the consistency decreases from the previous iterations. Note that unquantized observations are actually associated with 32-bits floating point variables.

For any fixed values of these parameters, 2000 trials of the MC simulations were considered by randomly drawing both the sparsity range profile $\bs a$, the radar sensing matrix $\bs \Phi$, and the dither $\bs \xi$. The resulting full resolution signals $\bs \Phi \bs a$ were then dithered and quantized from~\eqref{eq:ditheredquant} before estimation of $\bs a$ from PBP or QIHT.

More precisely, each $K$-sparse vector $\bs a$ was randomly built by picking its support uniformly at random among the $N \choose K$ possible supports, and by independently drawing its $K$ non-zero components as the random variable $C \exp(\im \psi)$, with $C \sim \cl U([0, 1])$ and $\psi \sim \cl U([0,2\pi))$, before the normalization $\bs a \leftarrow \bs a / \|\bs a\|_\infty$. Following Sec.~\ref{sec:radar-system-model}, the random sensing matrix $\bs \Phi = \bs F_\Omega^*$ was generated according to a random draw of $\cl T$ (inducing a random multiset $\Omega$). The complex dither $\bs \xi$ was generated as a complex random uniform vector adjusted to $\cl Q$ and $\delta = \alpha_b \Delta$ (see Sec.~\ref{sec:quantization}).  

We assessed the efficiency of the range profile estimation by measuring the accuracy of the support recovery. In particular, we computed the True Positive Rate, \ie ${\rm TPR}={\rm TP}/{K}$, where the number of True Positives $\rm TP$ is the number of estimated targets that were actually part of the true range profile, \ie $\rm TP := |\supp{\hat{\bs a}} \cap \supp{\bs a}|$.

As a first evaluation of the potential of dithered quantization, we have focused on the performances of PBP; hence establishing a reference level for further experiments. Fig.~\ref{fig:PBPtexK2} shows that for $\cl B = bM \in [2^8, ...,2^{13}]$, low bit-depth strategies (\eg $b\in {1,2}$) outperforms the TPR of high-resolution quantizers (with $M\leq N$ at $b=2^5$ in this bit-rate range). Moreover, in Fig.~\ref{fig:PBPtexK2} as well as in Fig.~\ref{fig:PBPtexK10}, we clearly observe a TPR saturation for undithered schemes from $M = \cl B/b \geq N$, \ie for $\cl B \geq 2^8$ and $\cl B \geq 2^9$ at one and two-bit quantization, respectively; deterministic quantization does not provide more information from repeated quantized measurements in our synthetic examples. The TPR performances of the dithered schemes, however, continue to scale as $\cl B$ increases, as hinted by~\eqref{PBPguarantee} and \eqref{eq:link-M-min-target-ampl-and-support-recov}. 

The simplicity of the PBP algorithm, \ie the absence of an explicit usage of the dither and the non-consistency of the produced estimate (see Sec.~\ref{sec:algo}), limits its ability to distinguish targets with weak amplitudes before the quantization level. Therefore, as observed by comparing the TPR of Fig.~\ref{fig:PBPtexK2} ($K=2$) and Fig.~\ref{fig:PBPtexK10} ($K=10$) for low resolutions dithered schemes, the PBP performances are rather poor for larger values of $K$ at identical values of $b$ and $M = \cl B/b$. We have thus compared the performances of QIHT --- which targets consistency and explicitly uses the dither --- and PBP in Fig.~\ref{fig:compare12bitK2} and Fig.~\ref{fig:compare12bitK10} in the context of 1-bit quantization, as well as in absence of quantization. In this last case, QIHT and PBP reduces to the IHT and Thresholding algorithms~\cite{BD2009,FR13}, respectively, and IHT also outperforms the Thresholding algorithm by fully exploiting the RIP of $\bs \Phi$~\cite{FR13}. In these two figures, the TPR of QIHT clearly exceeds the one of PBP in every quantization and bit-rate scenarios. Furthermore, for large values of $K$, the drop in performances in Fig.~\ref{fig:compare12bitK10} between the non-dithered and dithered schemes for the 1-bit PBP is reduced for 1-bit QIHT. In Fig.~\ref{fig:compare12bitK10}, the dithered $1$-bit QIHT is markedly better than any other methods for $\cl B = 2^{9}$ bits and above, reducing the bit-rate by as much as $93.75\%$ compared to the classic high resolution Nyquist sampling scheme. This bit-rate corresponds to $M\geq 2N = 512$ for 1-bit and $M\geq {N}/{2^{4}} = 16$ in absence of quantization; at harsh bit-rates quantity outweighs quality.
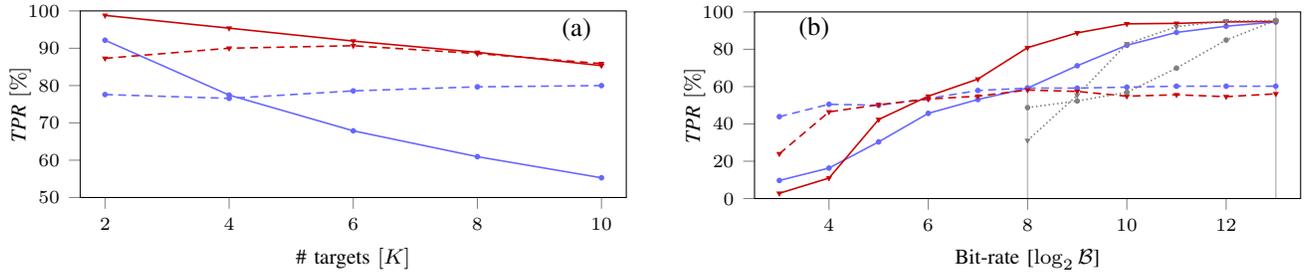
\begin{figure}[h]
\centering
\captionsetup[subfigure]{labelformat=empty}
\begin{subfigure}[b]{0.48\columnwidth}
% This file was created by matplotlib2tikz v0.6.17.
\begin{tikzpicture}

\definecolor{color0}{rgb}{0.117647058823529,0.564705882352941,1}
\def\sizep{0.75}
\begin{axis}[
width=1.\columnwidth,
height=\htex,
ylabel near ticks,
xlabel near ticks,
ytick={0,10,...,90,100},
xtick={0,2,...,10,12},
xlabel={\footnotesize \# targets $[K]$ },
ylabel={\footnotesize \textit{TPR} $[\%]$},
xmin=1.6, xmax=10.4,
ymin=50, ymax=100.,
tick align=outside,
tick pos=left,
ticklabel style = {font=\fontsize{7}{1}\selectfont},
ylabel shift = -8 pt,
x grid style={lightgray!92.02614379084967!black},
y grid style={lightgray!92.02614379084967!black}
]
%% Added by LJ, for including figure tag inside the figure, and save place 
\node at (800,450) {\fontsize{10}{1}\selectfont (a)};

\addplot [semithick, blue!60!white, mark=*, mark size=\sizep, mark options={solid}]
table {%
2 92.1503330161751
4 77.4500475737393
6 67.871868062163
8 60.941960038059
10 55.290199809705
};
\addplot [semithick, blue!60!white, densely dashed, mark=*, mark size=\sizep, mark options={solid}]
table {%
2 77.592768791627
4 76.569933396765
6 78.5759594037425
8 79.662226450999
10 79.9904852521408
};
\addplot [semithick, red!80!black, mark=triangle*, mark size=\sizep, mark options={solid,rotate=180}]
table {%
2 98.8106565176023
4 95.3853472882969
6 91.9283222327942
8 88.9034253092293
10 85.3472882968601
};
\addplot [semithick, red!80!black, densely dashed, mark=triangle*, mark size=\sizep, mark options={solid,rotate=180}]
table {%
2 87.2978116079924
4 90.0095147478592
6 90.6755470980019
8 88.5823025689819
10 85.8801141769743
};
\end{axis}

\end{tikzpicture}
\caption{}
\label{fig:support_PBP_QIHT}
\end{subfigure}
\begin{subfigure}[b]{0.48\columnwidth}
%\raisebox{1mm}{
% This file was created by matplotlib2tikz v0.6.17.
\begin{tikzpicture}

\definecolor{color1}{rgb}{0.415686274509804,0.352941176470588,0.803921568627451}
\definecolor{color0}{rgb}{0.117647058823529,0.564705882352941,1}
\definecolor{color3}{rgb}{0.862745098039216,0.0784313725490196,0.235294117647059}
\definecolor{color2}{rgb}{0.580392156862745,0,0.827450980392157}

\def\sizep{0.75}

\begin{axis}[
width=1.\columnwidth,
height=\htex,
xlabel={\footnotesize Bit-rate [$\log_2 \cl B$]},
ylabel={\footnotesize \textit{TPR} $[\%]$},
xmin=2.5, xmax=13.5,
ymin=0, ymax=100.,
xtick={0,2,...,12,14},
ytick={0,20,...,80,100},
ylabel near ticks,
xlabel near ticks,
x grid style={lightgray!92.02614379084967!black},
y grid style={lightgray!92.02614379084967!black},
ylabel shift = -8 pt,
tick align=outside,
tick pos=left,
ticklabel style = {font=\fontsize{7}{1}\selectfont}
]
\node at (12,91) {\fontsize{10}{1}\selectfont (b)};

\addplot [semithick,blue!60!white, mark=*, mark size=\sizep, mark options={solid}]
table {%
3 9.69387755102041
4 16.3265306122449
5 30.3571428571429
6 45.6632653061224
7 53.0612244897959
8 59.1836734693878
9 71.1734693877551
10 82.1428571428571
11 89.030612244898
12 92.3469387755102
13 94.6428571428571
};
\addplot [semithick, blue!60!white, densely dashed, mark=*, mark size=\sizep, mark options={solid}]
table {%
3 43.8775510204082
4 50.5102040816327
5 50
6 53.5714285714286
7 57.9081632653061
8 59.1836734693878
9 59.1836734693878
10 59.6938775510204
11 60.2040816326531
12 60.2040816326531
13 60.2040816326531
};
\addplot [semithick, black, nearly transparent, forget plot]
table {%
8 0
8 100
};
\addplot [semithick, black, nearly transparent, forget plot]
table {%
13 0
13 100
};
\addplot [semithick, red!80!black, mark=triangle*, mark size=\sizep, mark options={solid,rotate=180}]
table {%
3 2.80612244897959
4 10.969387755102
5 42.3469387755102
6 54.8469387755102
7 64.030612244898
8 80.8673469387755
9 88.7755102040816
10 93.6224489795918
11 93.8775510204082
12 94.6428571428571
13 94.8979591836735
};
\addplot [semithick, red!80!black, densely dashed, mark=triangle*, mark size=\sizep, mark options={solid,rotate=180}]
table {%
3 23.9795918367347
4 46.4285714285714
5 50.2551020408163
6 53.3163265306122
7 54.8469387755102
8 58.1632653061224
9 57.3979591836735
10 54.8469387755102
11 55.6122448979592
12 54.5918367346939
13 56.1224489795918
};

\addplot [semithick,lightgray!66.92810457516339!black, densely dotted, mark=*, mark size=\sizep, mark options={solid}]
table {%
8 48.7244897959184
9 52.2959183673469
10 56.8877551020408
11 69.8979591836735
12 84.9489795918367
13 95.4081632653061
14 nan
15 nan
16 nan
17 nan
18 nan
};

\addplot [semithick, lightgray!66.92810457516339!black, densely dotted, mark=triangle*, mark size=\sizep, mark options={solid,rotate=180}]
table {%
8 31.1224489795918
9 55.3571428571429
10 82.9081632653061
11 92.0918367346939
12 95.1530612244898
13 95.4081632653061
14 nan
15 nan
16 nan
17 nan
18 nan
};

\end{axis}

\end{tikzpicture}
%}
\caption{}
\label{fig:meas_AMG}
\end{subfigure}

\caption{[best viewed in color] (a) \textit{TPR} vs number of targets for $1$-bit PBP and $1$-bit QIHT with $\cl B = 2^9$ bits; (b) TPR vs bit-rate using real FMCW radar measurements for $K=2$. In all figures, PBP is represented by disks and QIHT by triangles, blue stands for $1$-bit PBP, red for $1$-bit QIHT, and gray for no quantization.}
\end{figure}

Finally, we study in Fig.~\ref{fig:support_PBP_QIHT} the TPR of PBP and QIHT vs $K$ for a fixed bit-rate $\cl B = 2^9$. Here also, the gain offered by the explicit knowledge of the dither in QIHT is quite obvious. For low $K$, both of the dithered schemes have better TPR than their non dithered counterparts. However, as $K$ increases above 4, the performances of the 1-bit dithered PBP plummets quickly below its non dithered version. On the other hand, QIHT with dithering always outperforms its performances with non dithered quantization. We thus conclude that, provided a uniform dithering can be implemented efficiently and later reproduced in QIHT, dithered quantization has always a positive impact on the range estimation.
\sq
\section{Measurement in Laboratory}
\label{sec:mes_radar}
Sec.~\ref{sec:simu} has focused on the study of range estimation performances from noiseless and synthetic simulations, under a perfect linear sensing model (before quantization) where an idealized radar interacts with point-like targets. We thus present here different tests of resilience of both our model and algorithms by confronting them with real data acquired in a controlled laboratory setting.  

The radar used for this experiment is the \textit{KMD2} radar~\cite{KMD2}, \ie an FMCW radar with one transmitting antenna and 3 receiving antennas. The radar lies in the ``K"-band and its bandwidth can be extended up to 770\,Mhz. The AMG43-007~\cite{AMG} is a target simulator distributed by AMG-microwave which is able to simulate a target with varying velocity, range, and power. In the context of this work, two simulators are used with the velocity set to zero and with a power changing according to a logarithmic uniform distribution. This setup allows to simulate target ranges up to 64\,m by 1\,m step. We thus set the bandwidth of the FMCW radar to 150\,Mhz to match this spatial resolution.

 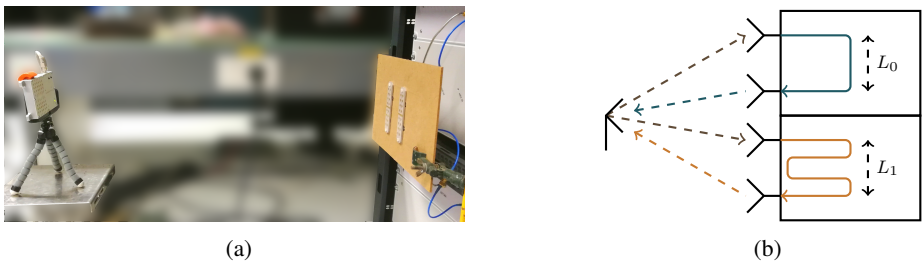
\begin{figure}[!t] 
    \centering
    \begin{subfigure}[b]{0.35\textwidth}
      \centering
      \scalebox{-1}[1]{
\includegraphics[width=0.95\columnwidth]{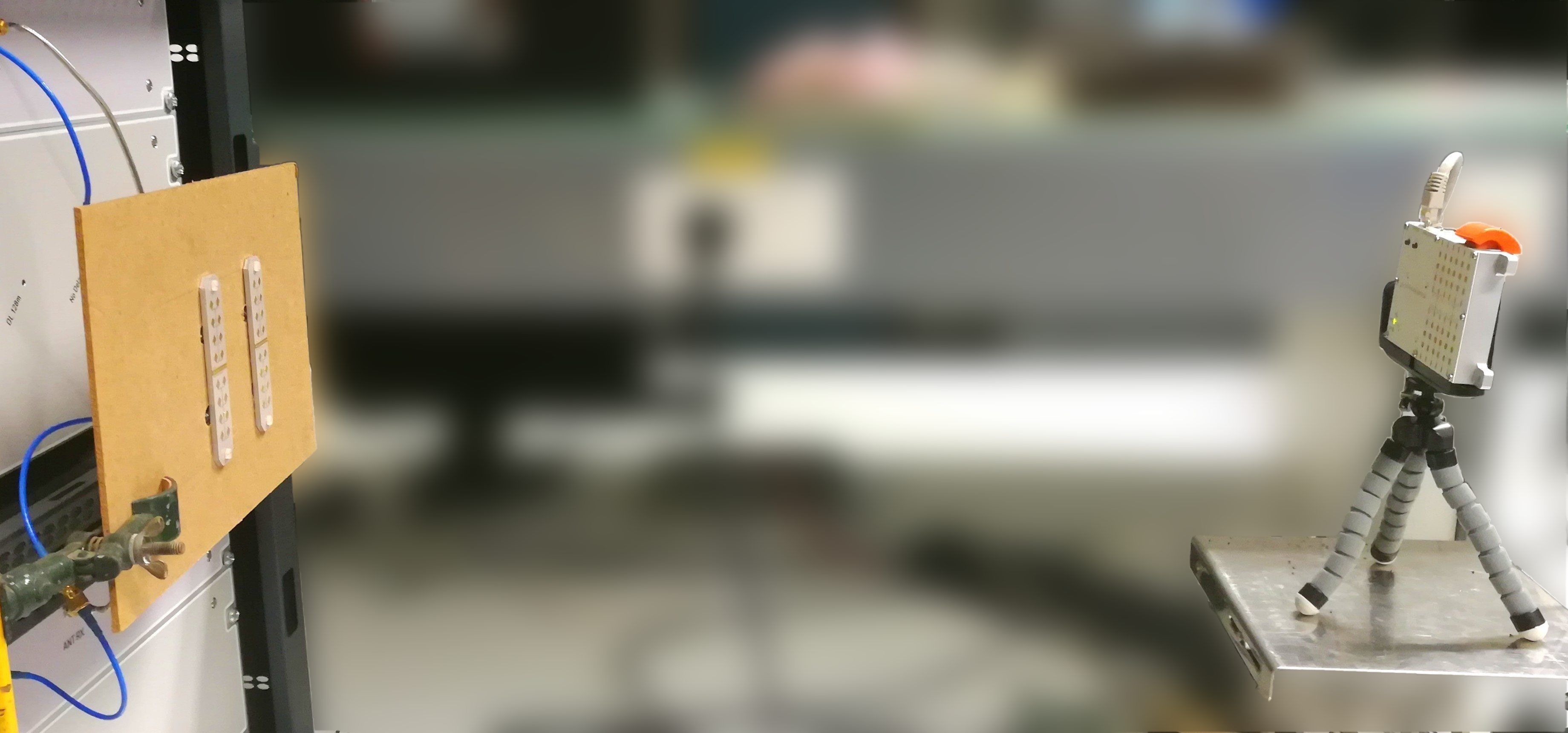}}
\caption{}
 \label{fig:AMGtot}
    \end{subfigure}
    \begin{subfigure}[b]{0.40\textwidth}
        \centering
\usetikzlibrary{through,calc}

\definecolor{bblue}{HTML}{4F81BD}
\definecolor{rred}{HTML}{C0504D}
\definecolor{ggreen}{HTML}{9BBB59}
\definecolor{ppurple}{HTML}{9F4C7C}

\definecolor{mDarkBrown}{HTML}{604c38}
\definecolor{mDarkTeal}{HTML}{23373b}
\definecolor{mMediumTeal}{HTML}{205A65}

\definecolor{mLightBrown}{HTML}{EB811B}
\definecolor{mMediumBrown}{HTML}{C87A2F}

\def\axesradargrid at (#1,#2){\draw[help lines,step=0.5] (0,0) grid (#1,#2);
\draw[->, thick](0,#2/2)--(#1,#2/2);
\node [right] at (#1,#2/2) {$\mathcal{R}$};
\draw[->, thick](#1/2,0)--(#1/2,#2);
\node [above] at (#1/2,#2) {$\mathcal{I}$};}

\def\axesradar at (#1,#2){
\draw[->, thick](0,#2/2)--(#1,#2/2);
\node [right] at (#1,#2/2) {$\mathcal{R}$};
\draw[->, thick](#1/2,0)--(#1/2,#2);
\node [above] at (#1/2,#2) {$\mathcal{I}$};}

\def\pointS at (#1:#2){
\draw [mDarkTeal,  thick,fill=mMediumTeal] (#1:#2) circle [radius=0.04];
\draw [mMediumBrown,  thick,dashed] (#1:#2) circle [radius=0.2];}

\def\dA{0.12}
\def\minihantenna at (#1,#2){\draw[thick] (#1,#2) --(#1-\dA,#2)--(#1-\dA-\dA,#2+\dA);
\draw[thick] (#1-\dA,#2) --(#1-\dA-\dA,#2-\dA);}

\begin{tikzpicture}[thick,scale=6.5/3.5, every node/.style={scale=3/3.5}]

\draw[ thick, dashed,->,mDarkBrown] (0.25,0.75)--(1.5-0.25,0.25/2+0.5+0.75-0.05);

\draw[ thick, dashed,->,mDarkBrown] (0.25,0.75)--(1.5-0.25,0.25/2+0.5-0.05);

%%%%%%%%%%%%%%%RX
\def\coefL{0.8}
\draw[ thick, dashed,<-,mMediumTeal] (1.25-\coefL,-\coefL*0.25/2-\coefL*0.05 +0.25/2+0.75+0.05)--(1.5-0.25,0.25/2+0.75+0.05);

\draw[ thick, dashed,<-,mMediumBrown] (1.25-\coefL,\coefL*0.75 -\coefL*0.25/2-\coefL*0.05+  0.25/2+0.05)-- (1.5-0.25,0.25/2+0.05);

%%Antenna

\draw[thick] (0.25,0.5) --(0.25,0.75) --(0.25+\dA,0.75+\dA);
\draw[thick] (0.25,0.75) --(0.25+\dA,0.75-\dA);

%%%%boxes
\draw[thick] (1.50,0) rectangle (2.5,0.75) ;
\draw[thick] (1.50,0.75) rectangle (2.5,1.5) ;

%% small antennas
\minihantenna at (1.5,0.25/2+0.05)
\minihantenna at (1.5,0.25/2+0.5-0.05)

\minihantenna at (1.5,0.25/2+0.75+0.05)
\minihantenna at (1.5,0.25/2+0.5+0.75-0.05)

%%%%%%%%%%%%%cable in it

\draw [->,rounded corners=2pt,  thick, mMediumTeal] (1.5,0.25/2+0.5+0.75-0.05) -- (1.5+0.5,0.25/2+0.5+0.75-0.05)-- (1.5+0.5,0.25/2+0.75+0.05) -- (1.5,0.25/2+0.75+0.05);

\draw [->,rounded corners=2pt,  thick, mMediumBrown] (1.5,0.25/2+0.5-0.05) -- (1.5+0.5,0.25/2+0.5-0.05)-- (1.5+0.5,0.25/2+0.5-0.05-0.25/2)--(1.5+0.5-0.45,0.25/2+0.5-0.05-0.25/2)--(1.5+0.5-0.45,0.25/2+0.05+0.25/2)--(1.5+0.5,0.25/2+0.05+0.25/2) --(1.5+0.5,0.25/2+0.05) -- (1.5,0.25/2+0.05);

%%%%%%%%%%%%%%%%%length
\draw[<->, thick,dashed] (2.125,0.25/2+0.05)--(2.125,0.25/2+0.5-0.05);
\node[right] at (2.125,0.25*3/2) {\footnotesize $L_1$};

\draw[<->, thick,dashed] (2.125,0.75+0.25/2+0.05)--(2.125,0.25/2+0.5-0.05+0.75);
\node[right] at (2.125,0.25*3/2+0.75) {\footnotesize $L_0$};

%%%%%%%%%%%%%%%%%%FinalLegend

\end{tikzpicture}
%\vspace{-.4cm}
\caption{}
        \label{fig:AMGtex}
    \end{subfigure}
 
   \caption{\ninept (a) Experimental setup: radar in front of the simulator. (b)~Block representation of the 2 targets simulator by \textit{AMG}}
    \label{AMG}
\end{figure}
The radar is placed in front of these two simulators and emits the frequency pattern~\eqref{txsig}, the signal received by the two simulators is then delayed and attenuated according to user defined parameters and then re-emitted towards the radar (see Fig.~\ref{fig:AMGtot} and Fig.~\ref{fig:AMGtex}). This process allows the simulation of a specific support while adding the concrete effect associated with the radar that are not taken into account in the developed model. These effects range from the inherent noise in RF and electronics hardware, IQ imbalance, non linearities in the coherent demodulation and all other non idealities related to radar applications. This experimental setup has thus the ability to combine the rigor and completeness of Monte Carlo simulations with the possibility to program and repeat specific scenarios (\ie specific $\bs a$), and to test them against a real acquisition system.

We recorded 196 runs with different sparse range profiles using the same parametrization as in Fig.~\ref{fig:compare12bitK2}. We observed that the SNR of the configuration in Fig.~\ref{fig:AMGtot} is sufficiently high to neglect the impact of the noise on the quantization. Note that this effect was briefly addressed in~\cite{COSERA} by experimentally adjusting the dither to the noise amplitude, and its thorough theoretical study is ongoing.

The curves in Fig.~\ref{fig:meas_AMG} exhibit the same tendencies than in Fig.~\ref{fig:compare12bitK2}. The only difference is the TPR at which the non dithered schemes saturate; an effect most probably due to some discrepancies in the range profile amplitudes between this setup and the previous simulations. Once again, 1-bit dithered QIHT is the algorithm with the highest TPR from $\cl B = 2^{6}$ bits to $2^{13}$, \ie the bit-rate of a full resolution Nyquist sensing. These results from real measurements are fully consistent with the previously developed theory and simulations; this paves the way to more complete and practical realizations of the proposed quantized architecture.
\sq
\section{Conclusion \& Future work}
\label{sec:conclusion}
In this article, we demonstrated that a pre-quantization dither removes unavoidable range estimation ambiguities when one quantizes the received radar signal. Moreover, in this dithered scheme, we proved that severe quantization, as low as 1-bit per received signal sample, still allows for an accurate range profile estimation as soon as the total bit-rate is large enough; a tradeoff between the number of radar observations (or measurements) and their resolution (or bit-depth) must be respected. Moreover, we showed that for low bit-rate scenarios, low bit-depth exhibits better performances than an unquantized scheme. These results are achieved thanks to two QCS reconstruction algorithms, PBP and QIHT, that leverage the sparsity of the range profile. Moreover, when the number of targets -- and thus the sparsity level of the range profile -- increases, Monte Carlo simulations proved that QIHT still provides high range estimation performances by promoting consistency with the quantized radar observations. As a proof of concept, we obtained similar range estimation performances from quantized observations of an actual radar in a controlled environment; hence showing that this QCS radar framework could apply in radar applications with limited bit-rate, \eg for radar signal reception with cheap ADC. Future work will address the interplay between the dither and the background noise, with a practical realization of the proposed highly quantized and dithered architecture. 

\ifCLASSOPTIONcaptionsoff
  \newpage
\fi


\begin{thebibliography}{1}

\bibitem{HS09} M. A. Herman, T. Strohmer, \emph{High-resolution radar via compressed sensing,} IEEE Trans. Signal Process.,  {\bf 57}(6):2275--2284, 2009.

\bibitem{End10} J. Ender, \emph{On compressive sensing applied to radar}, Signal Processing, {\bf 90}(5):1402--1414, 2010.

\bibitem{YONINA} D. Cohen, Y. Eldar, \emph{Sub-Nyquist Radar Systems: Temporal, Spectral and Spatial Compression}, IEEE Sig. Proc. Mag. (to appear). arXiv:1807.11542

\bibitem{BJSK15} P. T. Boufounos, L. Jacques, F. Krahmer, R. Saab, \emph{Quantization and compressive sensing}. In Compressed sensing and its applications, 2015 (pp. 193-237). Birkh\"auser, Cham.

\bibitem{FR13} S. Foucart, H. Rauhut, \emph{A mathematical introduction to compressive sensing}. {\bf 1}(3). Basel: Birkh\"auser, 2013.

\bibitem{candes} E. Cand\`es, J. Romberg, T. Tao, \emph{Stable signal recovery from incomplete and inaccurate measurements}, Commun. Pure Appl. Math., {\bf 59}(8):1207--1223, 2006.



\bibitem{MAP} X. Dong, Y. Zhang, \emph{A MAP approach for  1-bit Compressive  Sensing  in  Synthetic  Aperture  Radar  imaging}, IEEE Geo. Rem. Sens. Lett., {\bf 12}(6):1237--1241, 2015

\bibitem{varthres} J. Li, M. Naghsh, S. Zahabi, M. Modarres-Hashemi, \emph{Compressive radar sensing via one-bit sampling with time-varying thresholds}, In Sig., Sys. Comp., IEEE Asilomar Conf., 1164-1168, 2016.

\bibitem{2block} X. Wang, G. Li, Y. Liu, M. G. Amin, \emph{Enhanced 1-Bit Radar Imaging by Exploiting Two-Level Block Sparsity}, IEEE Trans. Geo.  Rem. Sens., 2018

\bibitem{COSERA} T. Feuillen, C. Xu, L. Vandendorpe,  L. Jacques, \emph{1-bit Localization Scheme for Radar using Dithered Quantized Compressed Sensing} CoSeRa 2018, arXiv:1806.05408

\bibitem{XJ2018} C. Xu, L. Jacques, \emph{Quantized compressive sensing with RIP matrices: The benefit of dithering}, arXiv:1801.05870, 2018.

\bibitem{ref1} R. M. Gray, T. G. Stockham, \emph{Dithered quantizers}, IEEE Trans. Inf. Th., {\bf 39}(3):805--812, 1993.

\bibitem{ref3} J. Rapp, R. M. A. Dawson, V. K. Goyal. \emph{Improving Lidar Depth Resolution with Dither}, IEEE Int. Conf. Im. Proc. (ICIP), 2018.

\bibitem{adcdither} B. Brannon, \emph{Overcoming Converter Nonlinearities with Dither}, Application note AN-410, ANALOG DEVICE, 1996

\bibitem{kevin} L. Jacques, K. Degraux, C. De Vleeschouwer,\emph{Quantized Iterative Hard Thresholding: Bridging 1-bit and High-Resolution Quantized Compressed Sensing}, Sampta 2013, arXiv:1305.1786

\bibitem{PV14} Y. Plan, R. Vershynin, \emph{Dimension  reduction  by  random  hyperplane  tessellations.} Discr. \& Comp. Geom., {\bf 51}(2):438--461, 2014.

\bibitem{ref2}E. H. Lloyd, \emph{Least-squares estimation of location and scale parameters using order statistics.} Biometrika {\bf 39}(1/2):88--95, 1952.

\bibitem{BD2009} T. Blumensath, M. Davies, \emph{Iterative Hard Thresholding for Compressed Sensing}, Appl. Comp. Harm. Anal., {\bf 27}(3):265--274, 2009.

\bibitem{JLBB13} L. Jacques, J. N. Laska, P. T. Boufounos, R. G. Baraniuk, \emph{Robust 1-bit compressive sensing via binary stable embeddings of sparse vectors}. IEEE Trans. Inf. Th., {\bf 59}(4):2082--102, 2013.

\bibitem{JL16} L. Jacques, \emph{Error decay of (almost) consistent signal estimations from quantized Gaussian random projections}. IEEE Trans. Inf. Th., {\bf 62}(8):4696--709, 2016.

\bibitem{KMD2} KMD2 radar transceiver, \url{https://www.rfbeam.ch/product?id=21}

\bibitem{AMG} AMG target simulator, AMG-043-007, \url{https://www.amg-microwave.com/-Test-et-Mesure-} 
%%%%%%%%%%%%%%%%%%%%%%%%%%%%%%%%%%%%%%%%%


\end{thebibliography}
\end{document}